\documentclass[11pt,twoside]{article}

\usepackage{asp2006}
\usepackage{epsf}
\usepackage{epsfig}
\usepackage{lscape}

\begin{document}
\title{Multicolor photometry of SU UMa and U Gem during quiescence, outburst and superoutburst.}
\author{P. Wychudzki, M. Miko\l{}ajewski, M. Wi\c{e}cek, A. Karska, C. Ga\l{}an, E. \'Swierczy\'nski, S. Fr\c{a}ckowiak, T. Tomov}

\affil{Nicolaus Copernicus University, Torun, Poland}

\begin{abstract}
The results of time-resolved observations of SU UMa and U Gem 
obtained  over two-years are presented. Both stars are prototypes of different classes 
of dwarf novae. We  studied brightness variations on different time scales: orbital,
QPO and flickering. The multicolor $BVRI$ photometry allows to distinguisch
the geometrical and physical sources of these variations.
\end{abstract}
\section*{Introduction}
The aim of our work is to separate geometrical phenomena (e.g. eclipses, 
disk precession) from physical phenomena which occur in two prototype dwarf
novae: SU UMa and U Gem. In particular, we study the 
influence
of such 
phenomena on flickering variability observed in our multicolor light curves
of these systems. The flashes are observed with the mean spacing 15 days
for the normal outbursts and about 260 days for superoutbursts in the case 
of SU UMa. Only normal outburst with mean spacing about 100 days are 
observed in the case of U Gem. Our observations were 
collected during three distinct phases of activity: quiescence, outburst
and superoutburst. For SU UMa, these are most probably the first  
observations obtained in all phases of activity using one 4-band photometric 
system. 

\section*{Observations}
We performed the observations from Jan 2006 to Feb 2008 using a 60 cm 
Cassegrain telescope at the Astronomical Observatory of the Nicolaus Copernicus
University in Piwnice, near Torun, Poland. We observed in $B$, $V$, $R_{\rm c}$ and
$I_{\rm c}$ bands and in a white light mode. In the current paper we do not present $V$ and $R_{\rm c}$
light curves.
 Exposure times for the SU UMa 
observations were 40 sec and 30 sec in $B$ and $V$ bands, respectively, and 
20-25 sec in $R_{\rm c}$ and $I_{\rm c}$ bands. The resultant time resolution was 
approximately 95 sec. The exposure times used for the U Gem observations were 20-60 
sec in $B$, 20-40 sec in $V$, 10-20 sec in $R_{\rm c}$ and 5-15 sec in $I_{\rm c}$, 
with time resolution of 135 sec. For the observations collected in the white light
mode for both stars, the exposure time was set to 5 sec which resulted in a time 
resolution of 8 sec. 
The level of accuracy of our measurements was 0.01 mag.
In order to phase our data, we adopted the following
ephemeris from the literature:\\

SU UMa: $HJD_{\rm ic}$ = 2446143.6627+0.076351 $\times$ $E_{\rm orb}$ (Thorstensen 1986)\\

U Gem: $HJD_{\rm min}$ = 2437638.8270+0.17690619 $\times$ $E_{\rm orb}$ (Krzemi\'nski 1965,
Marsh 1990). 
\\

Where $HJD_{\rm ic}$ concerns the spectroscopic inferior conjunction of
emission lines source for SU UMa, and $HJD_{\rm min}$ is the mideclipse
moment for U Gem.
Light curves of SU UMa obtained without filter and in $B$ and $I_{\rm c}$ bands
as well as $B-I_{\rm c}$ colour indicies 
are shown in Figure~1. The corresponding light and colour curves of $U$ 
Gem are presented in Figure 2.
We do not show our observations during outburst of U Gem because they do not
cover the whole orbital period.

\begin{figure}[!ht]
  \centering
  \epsfig{file=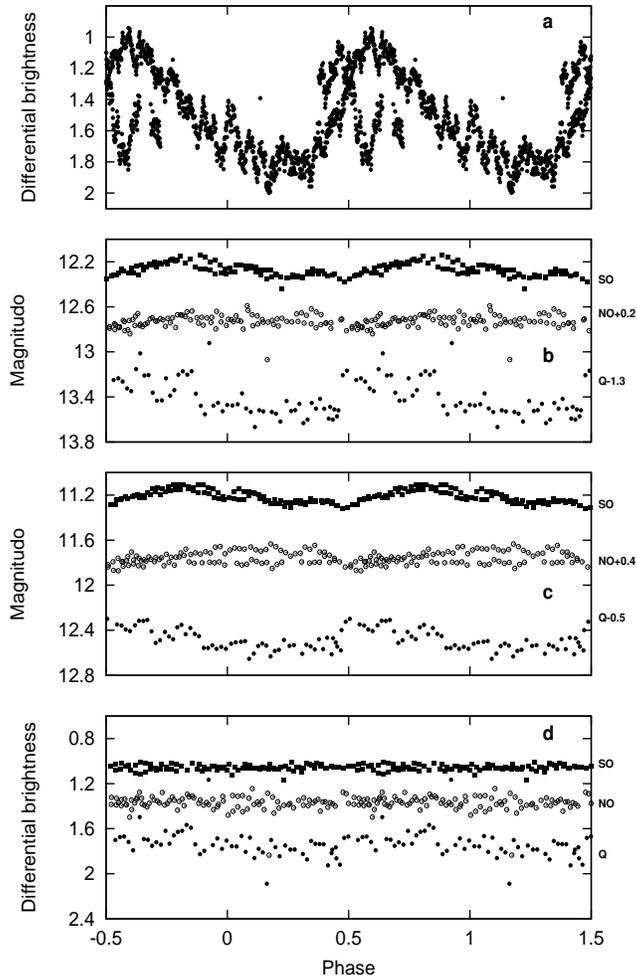,height=13.5cm,width=9cm}
 \caption{{\small 
 Light curves of SU UMa: (a) unfiltered at quiescence (Feb. 15, 2008); (b) $B$ light at quiescence (May 9, 2006), outburst (Sep. 12, 2006) and superoutburst (Sep. 25, 2006); (c) the same as previous pannel but in $I_{\rm c}$ light; (d) the $B-I_{\rm c}$ colour indicies obtained with the data from two upper pannels: quiescence (Q), outburst (NO), superoutburst (SO).}}
\end{figure}

\clearpage

\begin{figure}[!ht]
  \centering
  \epsfig{file=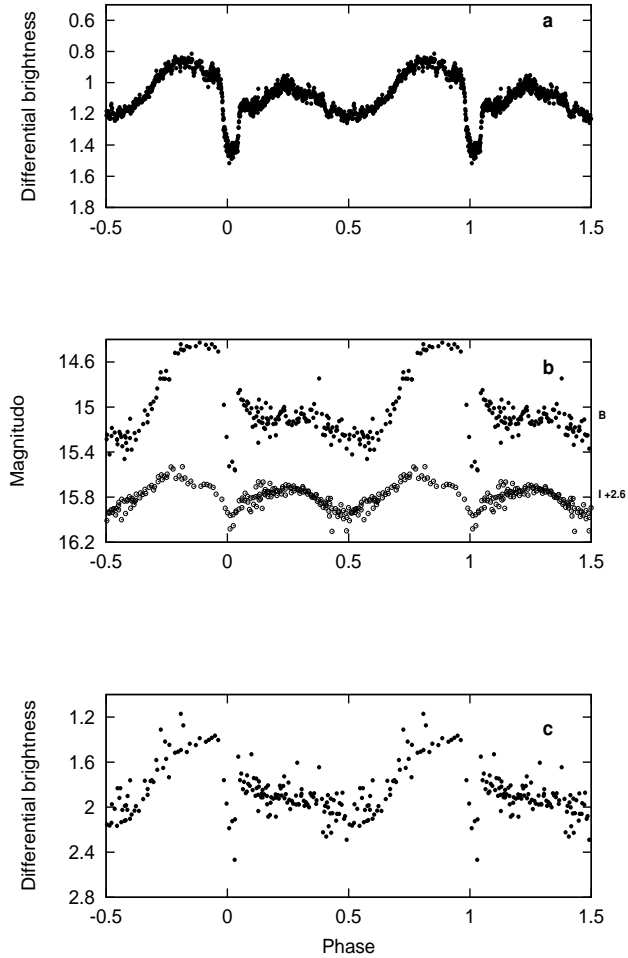,height=13.5cm,width=9cm}
 \caption{{\small 
Light curves of U Gem at quiescence: (a) unfiltered (28.12.2007); (b) $B$ and $I_{\rm c}$ filters (Jan. 7, Mar. 22, 23, May 9, 10, 14, 2006 ); (c) the $B-I_{\rm c}$ colour indices obtained with the data from previous pannel.}}
\end{figure}

\section*{Discussion}
We noticed fast flickering and orbital changes of the brightness during the quiescence phase in SU UMa 
 and U Gem systems (Figure 1 and 2). Especially both unfiltered
light curves show almost 
identical pattern of these variations during whole orbital period, if we neglect 
the eclipses in U Gem. The higher amplitude of such variations in the case of 
SU UMa is a result of the dominant influence of hot spot over the
accretion disk
and the other components of the system on the 
light curve. Moreover, the unfiltered
observations during the
quiescence and the $I_{\rm c}$ observations 
during the outburst phase, indicate an additional 
slow pattern of 
changes on timescales compatible with the orbital period.

%
%
During the superoutburst phase, characteristic superhumps are apparent in 
the light curves of SU UMa. $B-I_{\rm c}$ color index is constant for both 
superoutburst and outburst phases, which suggests that 
the changes of the system geometry are responsible for the observed 
superhumps. It was discussed by Udalski (1990) that the
superhumps are the result of
disk precession. 
The maxima of superhamps during superoutburst occur in different phases than
orbital maxima during quiescence.
The preccesion in the orbital
motion direction causes their period of changes to be longer than orbital period.

Highly time resolved observations in a non-filter mode in the quiescence 
show flickering variability on timescales of minutes and amplitudes
of 0.05-0.1 mag in U Gem (Figure 2a) and even 0.7 mag in SU UMa (Figure 1a). The flickering 
amplitude in both stars increases toward the shorter wavelenghts during the quiescence and almost 
disappear during the outburst and the superoutburst phases in SU UMa. The maximum 
amplitude corresponds to the best visibility of the hot spot (face-on) 
near phase 0.6. On the other hand, in the case of U Gem flickering is well visible 
not only when the hot spot is well seen (Figure~2a). The 
flickering is most pronounced near phase 0.4 when the hot spot does not face the 
observer. 
Thus, the influence of intrinsic accretion disk instabilities seem to be
a very important source of flickering in U Gem. \\

{ \bf Acknowledgements:} This work was supported by the Polish MNiSW Grant N203 018 32/2338.




\begin{thebibliography}{}

\bibitem[Krzeminski(1965)]{krzeminski65}Krzemi\'nski W., 1965, {\it ApJ}, {\bf 142}, 1051-1067
\bibitem[Marsh(1990)]{marsh90}Marsh T.R., Keith Horne 1990, {\it AJ}, {\bf 364},637-646
\bibitem[Thorstensen(1986)]{thorstensen86}Thorstensen J. R., Wade R. A., Oke J. B. 1986, {\it ApJ}, {\bf 309}, 721-731
\bibitem[Udalski(1990)]{udalski90}Udalski A., 1990, {\it AJ}, {\bf 100}, 26-232



\end{thebibliography}
\end{document}